\documentclass[prb,twocolumn,superscriptaddress,showpacs,floatfix,longbibliography]{revtex4-2}
\usepackage{tabularx}
\usepackage{amsmath}
\usepackage[margin=1in,letterpaper]{geometry}
\usepackage[final]{hyperref}
\hypersetup{
	colorlinks=true,
	linkcolor=blue,
	citecolor=blue,
	filecolor=magenta,
	urlcolor=blue
}
\usepackage{graphicx}
\usepackage{amssymb}
\usepackage{mathtools}
\usepackage{float,bm}
\usepackage{braket}

\begin{document}

\title{Optimal spectral transport of non-Hermitian systems}

\author{Mingtao Xu}
\affiliation{CAS Key Laboratory of Quantum Information, University of Science and Technology of China, Hefei 230026, China}
\author{Zongping Gong}
\email{gong@ap.t.u-tokyo.ac.jp}
\affiliation{Department of Applied Physics, The University of Tokyo, 7-3-1 Hongo, Bunkyo-ku, Tokyo 113-8656, Japan}
\author{Wei Yi}
\email{wyiz@ustc.edu.cn}
\affiliation{CAS Key Laboratory of Quantum Information, University of Science and Technology of China, Hefei 230026, China}
\affiliation{Anhui Province Key Laboratory of Quantum Network, University of Science and Technology of China, Hefei 230026, China}
\affiliation{CAS Center For Excellence in Quantum Information and Quantum Physics, Hefei 230026, China}
\affiliation{Hefei National Laboratory, University of Science and Technology of China, Hefei 230088, China}
\affiliation{Anhui Center for Fundamental Sciences in Theoretical Physics, University of Science and Technology of China, Hefei 230026 China}
\date{\today}

\begin{abstract}
The optimal transport problem seeks to minimize the total transportation cost between two distributions, thus providing a measure of distance between them. In this work, we study the
optimal transport of the eigenspectrum of one-dimensional non-Hermitian models as the spectrum deforms on the complex plane under a varying imaginary gauge field. Notably, according to the non-Bloch band theory, the deforming spectrum continuously connects the eigenspectra of the original non-Hermitian model (with vanishing gauge field) under different boundary conditions. It follows that the optimal spectral transport should contain key information of the model. Characterizing the optimal spectral transport through the Wasserstein metric, we show that, indeed, important features of the non-Hermitian model, such as the (auxiliary) generalized Brillouin zone, the non-Bloch exceptional point, and topological phase transition, can be determined from the Wasserstein-metric calculation. We confirm our conclusions using concrete examples.
Our work highlights the key role of spectral geometry in non-Hermitian physics, and offers a practical and convenient access to the properties of non-Hermitian models.
\end{abstract}

\maketitle

\section{Introduction}
The eigenstates of non-Hermitian systems are often sensitive to the boundary conditions~\cite{review1}. This gives rise to the now well-known non-Hermitian skin effect~\cite{skin1,skin2,skin3,skin4,skin5,skin6,skin7,skin8,skin9,skin10,skin11,skin12,skin13,skin14,acceleration1,acceleration2}, under which the nominal bulk eigenstates of the non-Hermitian Hamiltonian become localized toward open boundaries. At the same time, the geometries of the eigenspectra differ dramatically in the complex plane under different boundary conditions. Consider, for instance, one-dimensional (1D) non-Hermitian models with the non-Hermitian skin effect. Under the periodic boundary condition (PBC), the eigenpsectrum forms closed loops in the complex plane, with nontrivial spectral topology (protected by the so-called point gap) characterized by the spectral winding number. Whereas under the semi-infinite boundary condition, the spectrum fulfills the interior of the loop, occupying a finite area. These are contrasted with the case of open boundary condition (OBC), where the spectrum is reduced to tree-shaped arcs within the loop.
The drastic difference of eigenspectrum and eigenstates under different boundary conditions leads to the failure of the conventional bulk-boundary correspondence in non-Hermitian topological systems with skin effect~\cite{bbc1}. This is because the band gap (or the so-called line gap in the non-Hermitian case) closes under distinct parameters for different boundary conditions, rendering the bulk topological invariant, calculated using the extended Bloch states under the PBC, unable to account for the topological edge states under the OBC.
To restore the bulk-boundary correspondence, one can resort to the non-Bloch band theory~\cite{skin1,skin2,bbc2,amoeba,gbz1,gbz2,gbz3,gbz4,gbz5,gbz6}, which explicitly considers the localization of the bulk eigenstates under the OBC, by replacing the quasimomentum $k$ with $k-i\mu$, or equivalently, $e^{ik}$ with a complex variable $\beta=|\beta|e^{ik}$, with $|\beta|\neq 1$ in general. The non-Bloch topological invariant, which accounts for the topological edge states at open boundaries, can then be calculated over the generalized Brillouin zone (GBZ), defined through $\beta$ in the complex plane.
While the parameter $\mu$ can be considered as an imaginary flux through the ring formed by the 1D model under the PBC~\cite{hn}, its corresponding values for the GBZ, typically $k$-dependent, are determined by the eigenstates under the OBC~\cite{skin1}.
However, one can relax the GBZ condition and consider $\mu$ as a free variable.
As this imaginary flux varies, the eigenspectrum morphs through various geometries in the complex plane, continuously connecting eigenspectra of the original model (in the absence of the imaginary flux) under different boundary conditions, such as the PBC, the semi-infinite boundary condition, and the OBC.
Scanning $\mu$ or $\beta$ would also encompass all the auxiliary generalized Brillouin zones (aGBZs)~\cite{agbz1,agbz2}, which proves helpful for calculating the GBZ, predicting the presence of non-Bloch exceptional points (EPs)~\cite{ep1,ep2,pt1,pt2,pt3} and non-Hermitian dynamics~\cite{agbz3,agbz4}. Remarkably, in understanding the topological origin of the non-Hermitian skin effect, it has been proved that in 1D systems, by taking the intersection of the corresponding semi-infinite systems' eigenspectra, that is, interiors of the loops with winding number $w\neq 0$ under different $\mu$, one gets the eigenspectrum under the OBC~\cite{skin5,skin6}.
In higher dimensions, the recently discovered Amoeba formulation
of the non-Bloch band theory is also facilitated
by the concept of the $\mu$-space~\cite{amoeba}.
Given this context, and since the spectra of non-Hermitian systems contain much information regarding the system dynamics~\cite{phase}, two natural questions are whether one can establish a geometry in the $\mu$-space that quantifies the metamorphosis of the eigenspectrum as a function of $\mu$, and whether the geometric quantity provides a shortcut to key properties of the underlying non-Hermitian model.

In this work, we focus exclusively on 1D non-Hermitian systems, and characterize the morphing of the eigenspectrum using the Wasserstein metric~\cite{wasserstein1,wasserstein2,wasserstein3}, a quantity used for measuring the closest distance between different distributions. An alternative view is that we are studying the optimal spectral transport in the $\mu$-space.
As expected, we show that important features of the non-Hermitian model, such as the geometry of GBZ and the aGBZ,  the non-Bloch EP, and the topological phase transition, can be deduced from the Wasserstein-metric calculation.
We then illustrate our conclusions using the Hatano-Nelson (HN) model~\cite{hn,phase} and the non-reciprocal Su-Schrieffer-Heeger (SSH) model~\cite{skin1,semi1,semi2,semi3}. Furthermore, we use a non-Hermitian Aubry-Andr\'e (AA) model~\cite{quasi1,quasi2,quasi8,quasi9} to demonstrate that the Wasserstein metric can be used to determine the topological phase transition therein even in the absence of lattice translational symmetry where the notion of GBZ no longer applies~\cite{phase,quasi1,quasi2,quasi3,quasi4,quasi5,quasi6,quasi7,quasi8,quasi9,quasi10,quasi11,quasi12,LE1} .

The work is organized as follows. In Sec.~II, we review the Wasserstein distance and the Wasserstein metric, and apply them to describe the variation of the eigenspectrum in the complex plane. In Sec.~III, we discuss how properties of the non-Hermitian model are reflected in the Wasserstein metric. We also use HN and SSH models as examples to support our case. In Sec.~IV, we investigate quasiperiodic non-Hermitian models, where the Wasserstein metric can be applied to characterize the phase transitions therein. We summarize in Sec.~V.

\section{Wasserstein distance and Wasserstein metric}
In this section, we review the definitions of the Wasserstein distance and the Wasserstein metric in the context of the optimal transport problem~\cite{wasserstein1,wasserstein2,wasserstein3}. We then derive the expressions of the Wasserstein metric for 1D non-Hermitian models and discuss some of its key properties.

\subsection{Optimal transport, Wasserstein distance and metric}
In mathematics, the Wasserstein distance is introduced to solve the problem of optimal transportation~\cite{wasserstein1}.
Given a Euclidean space $X$  and a cost function $c(\boldsymbol{x},\boldsymbol{y})$ that represents the cost of transporting a particle from $\boldsymbol{x}\in X$ to $\boldsymbol{y}\in X$, the optimal transport problem, also known as the Monge-Kantorovich minimization problem, seeks to minimize the total transportation cost when the distribution $\rho(\boldsymbol{x})$ is changed (through transportation) to another $\sigma(\boldsymbol{y})$.
The Wasserstein distance between the two distributions is defined to be the minimized total cost.
Taking the cost function to be the $L^2$-norm $c(\boldsymbol{x},\boldsymbol{y})=||\boldsymbol{x}-\boldsymbol{y}||^2$, we define the
$L^2$-Wasserstein distance as
\begin{align}
        & (W\left[\rho(\boldsymbol{x}), \sigma(\boldsymbol{y})\right])^2 \notag\\
        & = \inf_{\pi\in\prod\left(\rho(\boldsymbol{x}), \sigma(\boldsymbol{y})\right)}\int_{X\times X}\left||\boldsymbol{x}-\boldsymbol{y}\right||^2\mathrm{d}\pi(\boldsymbol{x},\boldsymbol{y}),
\end{align}
where the lower bound $\pi$ is taken over the entire set of joint-probability distributions $\Pi\left(\rho(\boldsymbol{x}), \sigma(\boldsymbol{y})\right)$ on $X \times X$, whose marginal distributions reduce exactly to $\rho(\boldsymbol{x})$ and $\sigma(\boldsymbol{y})$. The $L^2$-Wasserstein distance is well-defined if the probability distributions satisfy
\begin{equation}
    \int \mathrm{d}\boldsymbol{x}\,\rho(\boldsymbol{x})||\boldsymbol{x}||^2<\infty,\quad \int \mathrm{d}\boldsymbol{y}\,\sigma(\boldsymbol{y})||\boldsymbol{y}||^2<\infty.
\end{equation}
In our work, these conditions are always satisfied.
Note that, while other distance measures (such as the Kullback-Leibler divergence~\cite{amari}, which is, however, ill-defined is our setting) exist, the Wasserstein distance is a better fit here for its appropriate properties and broader applications.

The continuous formalism above has a discrete extension, which we adopt in the following sections to characterize finite-size systems.
We define the optimal transport matrix $T$, which is a matrix representation of the transport. While the entries of the matrix $T$ take values from $\{0,1\}$, a unit matrix element $T_{ij}$ indicates transporting a source point $\boldsymbol{x_i}\in X$ to a target point $\boldsymbol{y_j}\in X$. We also define the cost matrix $C$, whose element $C_{ij}=c(\boldsymbol{x_i},\boldsymbol{y_j})=||\boldsymbol{x_i}-\boldsymbol{y_j}||^2$ represents the cost
of the transportation $T_{ij}$.
The optimal transport problem is to find a matrix $T$ to minimize $\sum_{ij}T_{ij}C_{ij}$ and the corresponding Wasserstein distance is $W^2=\frac{1}{N}\min_T\sum_{ij}T_{ij}C_{ij}$.

So far we have talked about the Wasserstein distance on the whole (infinite dimensional) space of probability distributions. In practical applications, we usually assume a statistical model $\rho(\boldsymbol{x};\boldsymbol{\theta})$ with a finite dimensional parameter space of $\boldsymbol{\theta}$. To further obtain the Wasserstein metric induced on the parameter space, we could follow a procedure similar to that used in the derivation of the Fisher information metric~\cite{wasserstein3}. However, this approach can be challenging to apply, since it requires analytic solution of the score function defined in information geometry, which further requires analytic solution of the model. A more straightforward method for deriving the Wasserstein metric is to start from its definition
 \begin{align}
        & \left(\mathrm{d}W[\rho(\boldsymbol{\theta}), \rho(\boldsymbol{\theta}+\mathrm{d}\boldsymbol{\theta})]\right)^2  \notag \\
        & = \inf_{\pi\in\prod\left(\rho(\boldsymbol{\theta}), \rho(\boldsymbol{\theta}+\mathrm{d}\boldsymbol{\theta})\right)}\int_{X\times X}||\boldsymbol{x}-\boldsymbol{y}||^2 \mathrm{d}\pi(\boldsymbol{x},\boldsymbol{y}) \notag \\
         & = \mathrm{d}\boldsymbol{\theta}^T G_W(\boldsymbol{\theta}) \mathrm{d}\boldsymbol{\theta},
\end{align}
where $G_W$ is the Wasserstein metric.

\subsection{Wasserstein metric for 1D non-Hermitian models}
In this section, we derive the Wasserstein metric on the $\mu$ space for generic 1D non-Hermitian models.
We start with a single-band model, whose (non-)Bloch Hamiltonian can be generally written as
$H(\beta)=\sum_{n=-p}^q t_n \beta^n$, where $p$ and $q$ represent the left and right hopping ranges, respectively.
Note that, in general, the spectrum of a 1D non-Hermitian Hamiltonian under PBC forms closed loops in the complex energy plane, and may exhibit self-crossing points. In the following, we first derive the Wasserstein metric in the simpler case, where the eigenspectra do not self intersect.

In the following, we denote the eigenspectrum of $H(e^{ik+\mu})$ as $E(k,\mu)$, which corresponds to a simple closed curve in the complex plane with no self-crossing point. Under a small change $\Delta\mu$ in the parameter space, we analyze the Wasserstein distance between the spectra $E(k_i, \mu + \Delta\mu/2)$ and $E(k_j, \mu-\Delta\mu/2)$. Notably, we have $i\frac{\partial E}{\partial \mu}=\frac{\partial E}{\partial k}$.
Therefore, the Wasserstein metric in the thermodynamic (which is taken after first taking the continuous limit $\Delta\mu\to 0$) can be calculated analytically as follows
\begin{align}
    & G_W(\mu) \notag=\lim_{N\to\infty,,\Delta\mu\to 0} \min_T\frac{1}{N}\notag \\
    & \sum_{k_i,k_j}T_{ij}\left| \frac{E(k_i, \mu + \frac{\Delta\mu}{2}) - E(k_j, \mu - \frac{\Delta\mu}{2})}{\Delta\mu} \right|^2 \notag \\
    & =\lim_{N\to\infty,\Delta\mu\to 0} \min_T \frac{1}{N} \notag \\
    & \sum_{k_i,k_j}T_{ij}\left| \frac{E(k_i, \mu) - E(k_j, \mu) + \frac{\partial E}{\partial \mu}(k_i, \mu)\Delta\mu}{\Delta\mu}\right|^2 \notag \\
    & =\lim_{N\to\infty} \min_T\frac{1}{N} \notag \\
    & \sum_{k_i,k_j}T_{ij}\left\{\left| \frac{\partial E}{\partial k}(k_i, \mu)\frac{\Delta k_{ij}}{\Delta\mu} \right|^2 + \left|\frac{\partial E}{\partial \mu}(k_i, \mu)\right|^2 \right\} \notag \\
    & = \lim_{N\to\infty} \min_T\frac{1}{N} \sum_{k_i,k_j} T_{ij}\left|\frac{\partial E}{\partial \mu}(k_i, \mu)\right|^2\notag \\
    & =\lim_{N\to\infty} \frac{1}{N}\sum_{k_i}\left|\frac{\partial E}{\partial \mu}(k_i, \mu)\right|^2,
\end{align}
where we assume that $k_j$ is close to $k_i$, and define $\Delta k_{ij}=k_i-k_j$.
This result indicates that, in the absence of spectral self-crossings, the Wasserstein distance between the spectra  $E(k_i,\mu+\Delta\mu/2)$ and $E(k_j, \mu-\Delta\mu/2)$ is simply the sum of the Euclidean distances between the corresponding wave numbers $k$. The corresponding Wasserstein metric is then
    \begin{align}
            G_W(\mu) & = \frac{1}{2\pi}\int_{0}^{2\pi}\left|\frac{\partial E}{\partial \mu}(k, \mu)\right|^2\mathrm{d} k  \\
           & = \frac{1}{2\pi}\int_{0}^{2\pi}\left|\frac{\partial E}{\partial k}(k, \mu)\right|^2\mathrm{d} k. \label{1dsingleband}
    \end{align}

In the case of spectral self crossing, the situation is more complicated, especially for a finite-size system. Then the Wasserstein metric contains a finite-size contribution if and only if the self-crossing point corresponds to spectral degeneracy.

\begin{figure}[tbp]
    \centering
    \includegraphics[width=1\linewidth]{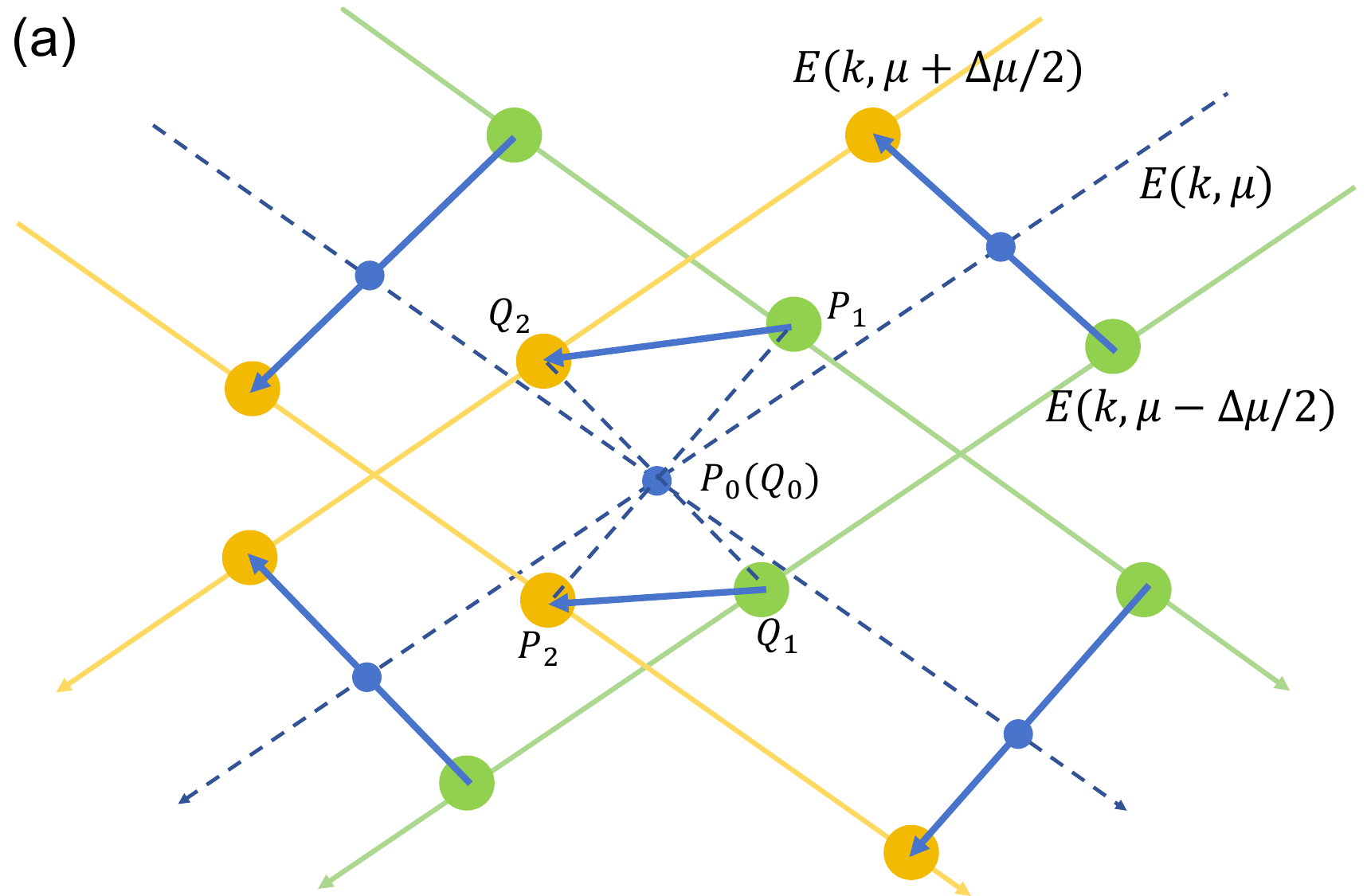}
    \vspace{1mm}

    \includegraphics[width=1\linewidth]{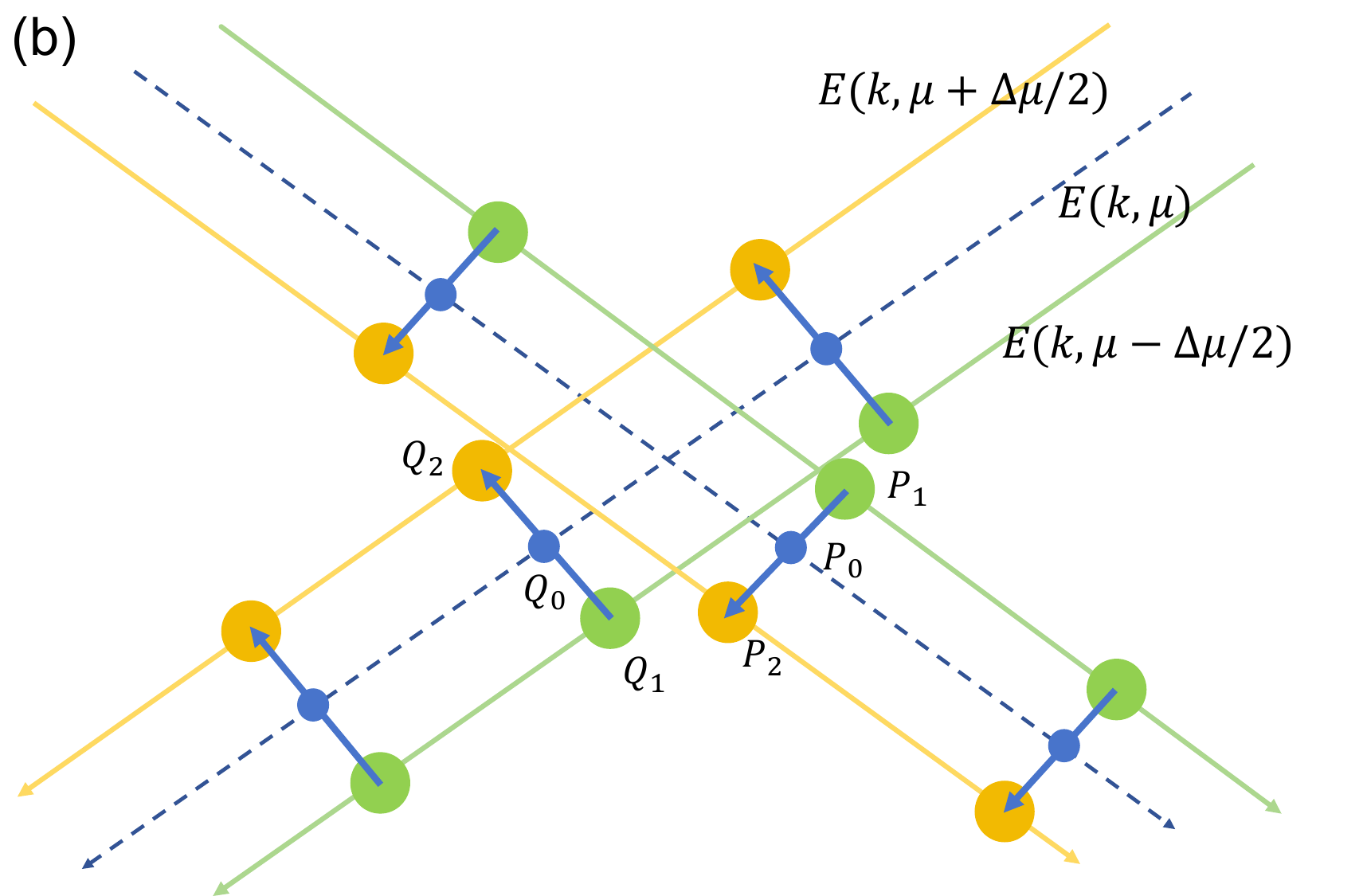}

    \caption{Schematic illustration of the optical transport close to a spectral self-crossing point, where we use solid blue lines to describe the optimal transport. (a)
     The case where the self-crossing point corresponds to spectral degeneracy $P_0=Q_0$. Abnormal transport occurs close to the self-crossing point, with either $P_1\to Q_2$ or $P_2\to Q_1$.
     (b) The case where there is no spectral degeneracy. We can always find a small enough $\Delta\mu$ to ensure the optimal transport between the same $k$.}
    \label{Fig1}
\end{figure}

First, as illustrated in Fig.~\ref{Fig1}(a), we denote the self-crossing point as $P_0$ ($Q_0$) and the corresponding wave number as $k'$ ($k''$), so that the spectrum is degenerate at $E(k',\mu)= E(k'',\mu)$ (or, equivalently, $P_0=Q_0$).
With the optimal transport in mind, we denote
$P_1$ ($Q_1$) and $P_2$ ($Q_2$) as the corresponding points on the morphed eigenspectra, respectively, with $P_{1,2}=E(k',\mu\mp\Delta\mu/2)$ and $Q_{1,2}=E(k'',\mu\mp\Delta\mu/2)$.
Since $P_1Q_1P_2Q_2$ is a trapezoid, we have the inequality
\begin{equation}\label{inequality}
    |P_1Q_2|^2+|P_2Q_1|^2\leq |P_1P_2|^2+|Q_1Q_2|^2,
\end{equation}
where the equality holds only when $P_1$ coincides with $Q_1$ and $P_2$ coincides with $Q_2$.
Thus, in a finite-size system, there always exists an abnormal optimal transport
in the neighbourhood of the crossing to further minimize the Wasserstein distance.
In the case of Fig.~\ref{Fig1}(a), the abnormal transport corresponds to either $P_1\to Q_2$ or $P_2\to Q_1$, instead of $P_1\to P_2$ and $Q_1\to Q_2$.
Furthermore, since the Wasserstein distance between $P_1,Q_2$ and $P_2,Q_1$ is proportional to $\Delta\mu$, this abnormal contribution does manifest itself in the Wasserstein metric when taking the limit of $\Delta\mu\to0$.

Second, due to the discrete nature of finite-size systems, spectral degeneracy may not arise even in the presence of spectral self crossing. This is illustrated in Fig.~\ref{Fig1}(b), where $P_0$ and $Q_0$ denote the eigenvalues closest to the crossing point.
Unlike before, in this case, we can always find a small enough $\Delta\mu$ to make the optimal transport normal, that is, $P_1\to P_2$ and $Q_1\to Q_2$.

To quantify the effect discussed above, for simplicity, we consider the scenario where there is only one
spectral degeneracy point $E(k',\mu)=E(k'',\mu)$, under the parameter $\mu$. The Wasserstein metric can be expressed as the sum of two terms
    \begin{align}\label{cross}
        & G_W(\mu) = \frac{1}{N}\sum_{k_i\neq k',k''}\left|\frac{\partial E}{\partial \mu}(k_i, \mu)\right|^2 \notag \\
        & + \lim_{\mathrm{\Delta}\mu\to0}\frac{1}{N} \quad\left| \frac{E(k'', \mu + \frac{\Delta\mu}{2}) - E(k', \mu- \frac{\Delta\mu}{2})}{\Delta\mu} \right|^2 \notag \\
        & + \lim_{\mathrm{\Delta}\mu\to0}\frac{1}{N} \quad\left| \frac{E(k', \mu + \frac{\Delta\mu}{2}) - E(k'', \mu- \frac{\Delta\mu}{2})}{\Delta\mu} \right|^2 \notag \\
        &  = \frac{1}{N}\sum_{k_i}\left|\frac{\partial E}{\partial \mu}(k_i, \mu)\right|^2-\frac{2}{N}\left|\frac{\partial E(k',\mu)}{\partial \mu}-\frac{\partial E(k'',\mu)}{\partial \mu}\right|^2
    \end{align}
where in the first term the optical transport occurs between the same $k$. The second term arises from the finite-size effect discussed above.

In the thermodynamic limit $N\rightarrow \infty$, the second term vanishes, and the expression for the Wasserstein metric is the same as Eq.~(\ref{1dsingleband}).
However, for finite-size systems, the contribution from the second term in Eq.~(\ref{cross}) is non-vanishing in the presence of exact degeneracies, as discussed above.
Therein, the contribution from the second term is of the order $\mathcal{O}\left(1/N\right)$, and
makes the Wasserstein metric smaller at a discrete set of $\mu$ (see later discussions). This serves as a basis for determining the GBZ and aGBZ from the metric calculations.

We now move on to the multi-band models. Based on the discussions above, the Wasserstein metric for multi-band systems can be expressed as $G_W(\mu) = \sum_{i=1}^m\frac{1}{2\pi}\int_{0}^{2\pi}\left|\frac{\partial E_i}{\partial k}(k, \mu)\right|^2\mathrm{d} k$, where the subscript $i$ is the band index and $m$ is the total number of bands. For the convenience of later discussions,
in the following, we derive an alternative expression for the multi-band Wasserstein metric in terms of the (non-)Bloch Hamiltonian and its right eigenvectors.

We first rewrite the eigenspectrum as $E(k,\mu)=u(k,\mu)+iv(k,\mu)$, where $u$ and $v$ denote the real and imaginary parts of the spectrum, respectively. It is important to note that the complex  spectrum behaves as a harmonic function in the $k-\mu$ space. Thus, we have the relations: $\frac{\partial u}{\partial \mu} = \frac{\partial v}{\partial k}$ and $\frac{\partial u}{\partial k} = -\frac{\partial v}{\partial \mu}$. We find that the area enclosed by the spectrum, given by $\mathcal{A}(\mu) = \int_0^{2\pi}u\partial_k v\mathrm{d}k$, is related to the Wasserstein metric. Namely,
\begin{equation}\label{eq:area}
        \frac{\mathrm{d}\mathcal{A}}{\mathrm{d}\mu} = \int_0^{2\pi}\left[(\partial_ku)^2+(\partial_kv)^2\right]\mathrm{d}k=2\pi G_W.
\end{equation}
Note that, by definition, the area $\mathcal{A}(\mu)$ acquires a sign depending on the circulation direction of the PBC energy loop (or, equivalently, the spectral winding number). Physically, in 1D non-Hermitian models, the spectral area $\mathcal{A}(\mu)$ correspond to the transient self acceleration of local excitations~\cite{acceleration1,acceleration2}.
This provides us with another geometric interpretation of the Wasserstein metric, indicating that it is proportional to the rate of change of the area enclosed by the non-Hermitian eigenspectrum as the parameter $\mu$ varies. Noting that in multi-band systems, $\mathcal{A}(\mu)= \frac{1}{4}\frac{\mathrm{d}}{\mathrm{d}\mu}\int_0^{2\pi}\sum_{i=1}^m|E_i|^2\mathrm{d}k$, we have
    \begin{align}\label{multibands}
     G_W(\mu) &  = \frac{1}{2\pi}\frac{\mathrm{d}\mathcal{A}}{\mathrm{d}\mu}
         = \frac{1}{8\pi}\frac{\mathrm{d}^2}{\mathrm{d}\mu^2}\int_0^{2\pi}\sum_{i=1}^m|E_i|^2\mathrm{d}k \notag\\
        & = \frac{1}{8\pi}\frac{\mathrm{d}^2}{\mathrm{d}\mu^2}\int_0^{2\pi}\mathrm{Tr}(HMH^\dagger M^{-1})\mathrm{d}k,
    \end{align}
where $M=U^\dagger U$, and each column of $U$ corresponds to a right eigenvector of the non-Hermitian Hamiltonian $H$~\cite{review1}.
Note that the multi-band formula Eq.~(\ref{multibands}) is consistent with $G_W(\mu) = \sum_{i=1}^m\frac{1}{2\pi}\int_{0}^{2\pi}\left|\frac{\partial E_i}{\partial k}(k, \mu)\right|^2\mathrm{d} k$.

\subsection{Key Properties}
Before ending the section, we would like to highlight some mathematical properties of the Wasserstein metric.

First, the Wasserstein metric for a multi-band model can have singularities. A singularity occurs for $\beta_c = e^{ik_c+\mu_c}$, if and only if $H(\beta_c)$ is at an EP.
To prove this, we first make the following expansion in the vicinity of the EP: $E_i(k_c + \delta k, \mu_c)  \simeq E(k_c,\mu_c) + c_i(\delta k)^{1/n}$ ($n$ is the number of coalescing bands). This leads to
\begin{equation}\label{EP}
            \left|\frac{\partial E_i}{\partial  k}\right|^2 \Bigg|_{k_c + \delta k} \simeq \left|\frac{c_i^2}{n^2(\delta k)^{2-\frac{2}{n}}}\right|.
\end{equation}
The divergence occurs for $n\geq2$, giving rise to a singularity in the integral and the metric.
On the other hand, according to Eq.~(\ref{multibands}), the Wasserstein metric diverges only when the matrix $M$ is ill-defined, which occurs when the non-Hermitian Hamiltonian is non-diagonalizable, that is, at an EP of the Hamiltonian.

Second, the Wasserstein metric in the thermodynamic limit is a convex function in the absence of singularities. To simplify analysis, we use the single-band formula Eq.~(\ref{1dsingleband}), which
gives
\begin{equation}\label{convex}
         \frac{\mathrm{d}^2 G_W}{\mathrm{d}\mu^2}
                 = \frac{2}{\pi}\int_{0}^{2\pi}\left\{\left(\frac{\partial^2 u}{\partial k^2}\right)^2 + \left(\frac{\partial^2 v}{\partial k^2}\right)^2\right\}\mathrm{d}k.
\end{equation}
Hence $\frac{\mathrm{d}^2 G_W}{\mathrm{d}\mu^2}\geq 0$.
Based on Eq.~(\ref{convex}) and the multi-band expression for the Wasserstein metric, it is straightforward to show that in the multi-band case, we also have $\frac{\mathrm{d}^2 G_W}{\mathrm{d}\mu^2}\geq 0$ in the absence of singularities, or in each interval separated by singularities.

\section{Spectral features through the Wasserstein metric}
In this section, we show how one may use the Wasserstein metric to infer key features of a non-Hermitian model.

We first give a brief review of the non-Bloch band theory for 1D non-Hermitian systems~\cite{skin1,skin2}. We consider a single-band model with the Bloch Hamiltonian $H(k)=\sum_{n=-p}^q t_n e^{ink}$ under the PBC. Following the non-Bloch band theory, we introduce an imaginary flux $\mu$ as a function of $k$, and write the non-Bloch Hamiltonian as $H(\beta)=\sum_{n=-p}^q t_n \beta^n$, with $\beta=e^{ik+\mu}$. The characteristic equation of the Hamiltonian is

\begin{align}
f(E,\beta) & = \det(E-H(\beta)) = E - \sum_{n=-p}^q t_n \beta^n.
\end{align}

Sorting its roots by their module, we have $|\beta_1(E)|\leq|\beta_2(E)|\leq\cdots\leq|\beta_{p+q}(E)|$. In the thermodynamic limit, the OBC results in $|\beta_p(E)|=|\beta_{p+1}(E)|$ (the GBZ condition)~\cite{gbz2,amoeba}, whose solution is the GBZ. The GBZ forms a closed curve in the complex $\beta$-plane.
To facilitate the calculation of the GBZ, the aGBZs are introduced in Ref.~\cite{agbz1}, where the $n$ th aGBZ satisfies the condition $|\beta_n(E)|=|\beta_{n+1}(E)|$ (the aGBZ condition), with $n=1,2,\cdots,p+q-1$.

\subsection{GBZ and aGBZ}

\begin{figure*}[tbp]
    \centering
    \includegraphics[width=\linewidth]{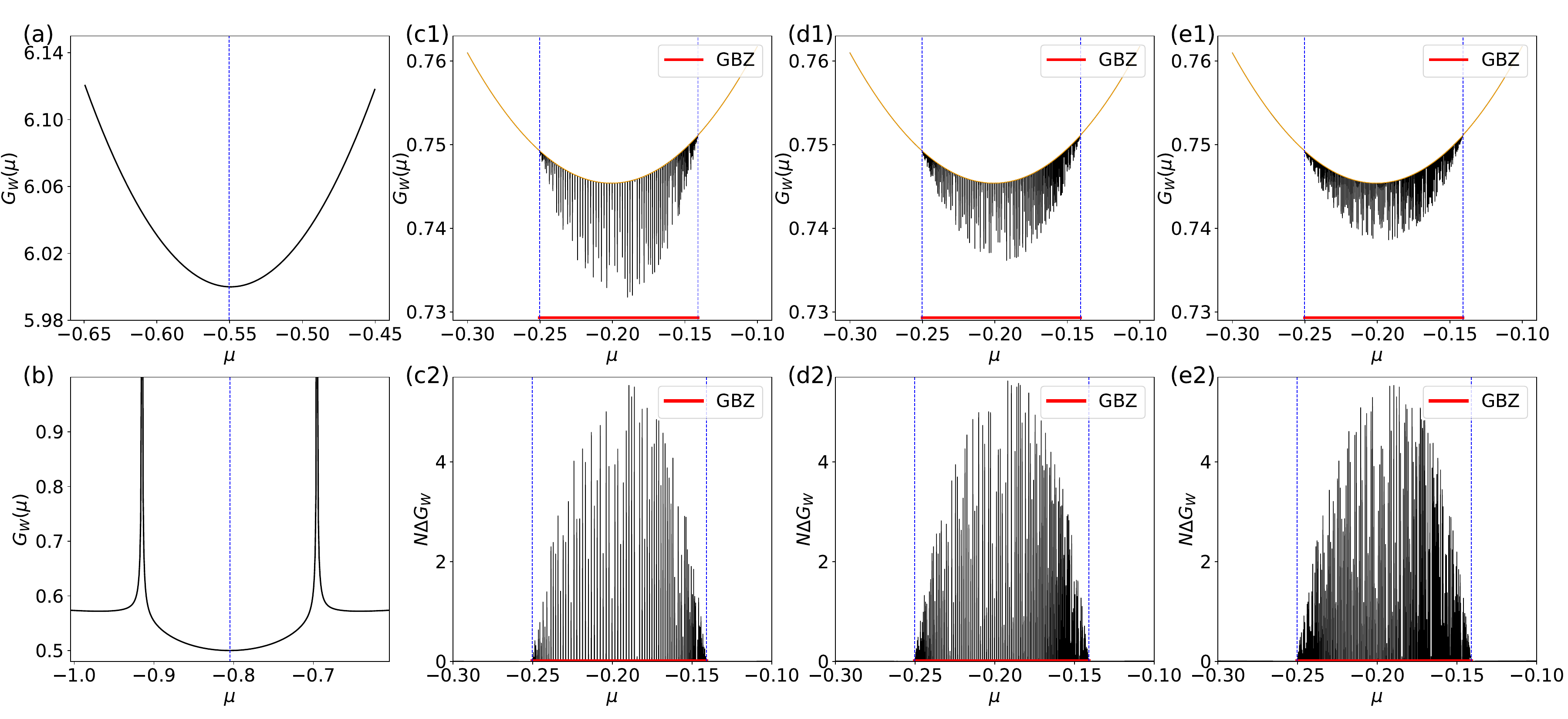}
    \caption{Determining the modulus range of GBZ. (a) The Wasserstein metric for the HN model with hopping parameters $(t_L,t_R)=(3,1)$. The GBZ is circular, with a radius $\mu_{\text{GBZ}}=\mu_{\text{min}}=\frac{1}{2}\ln\frac{t_R}{t_L}\simeq-0.549$. (b) The Wasserstein metric for a non-reciprocal SSH model with hopping parameters $(t_1,t_2,t_3,\gamma)=(1.3,1,0,4/3)$.
    The GBZ is also circular, with $\mu_{\text{GBZ}}$ corresponding to the minimum between the singularities.  (c1)-(e1) $G_W$ and (c2)-(e2) $N\Delta G_W$ as functions of $\mu$ for the non-reciprocal SSH model, with $\Delta\mu=10^{-4}$. Here a sufficiently small $\Delta \mu$ should be chosen, to ensure $\Delta\mu<\Delta k=2\pi/N$ and numerical convergence.
     The system size is $N=200,300,400$ for (c1)(c2), (d1)(d2), and (e1)(e2), respectively.
    The hopping parameters are $(t_1,t_2,t_3,\gamma)=(2.5,1,0.2,4/3)$. The GBZ is non-circular in this case, and its modulus range is indicated by the solid red line on the $\mu$-axis, where finite-size effects are observed.
   }
    \label{Fig2}
\end{figure*}
In this subsection, we study the relation between the Wasserstein metric, the GBZ, and the aGBZ.

We first consider a special case where the GBZ forms a circle in the complex $\beta$-plane, with a finite radius \(|\beta_{\text{GBZ}}| = e^{\mu_{\text{GBZ}}}\). 
For a large class of models with a constant $\mu_{\text{GBZ}}$, the PBC spectra for $\mu_{\text{GBZ}}\pm\mu$ are identical. This is the case for all the models that we consider in this section.
According to the definition of Wasserstein metric, $G_W(\mu) = \sum_{i=1}^m\frac{1}{2\pi}\int_{0}^{2\pi}\left|\frac{\partial E_i}{\partial k}(k, \mu)\right|^2\mathrm{d} k$, the metrics for $\mu_{\text{GBZ}}\pm\mu$ are identical, too. Since the Wasserstein metric is a convex function, the axis of symmetry \(\mu_{\text{GBZ}}\) aligns with the local minimum $\mu_{\text{min}}$. In other words, \(\mu_{\text{min}} = \mu_{\text{GBZ}}\). In multi-band models, because of the presence of singularities, there may be many local minima, and only the central one corresponds to the radius of the GBZ (see the corresponding discussions below).

On the other hand, for the extreme case where $\mu_{\text{GBZ}}=\pm\infty$ (which arises when
the system features unidirectional hopping for instance), the energy spectrum collapses to a point rather than an arc. In this case, the Wasserstein metric still takes its minimum at $\mu_{\text{GBZ}}$ but $\mu_{\text{GBZ}}=\pm\infty$ is no longer the axis of symmetry.

For the general case of non-circular GBZs, it is necessary to first understand the relation between the aGBZ condition and spectral self-crossing~\cite{agbz1,agbz2}.
For simplicity, we use single-band models as an example. While the characteristic equation is $\det(E-H(\beta))=E-H(\beta)=0$, the aGBZ condition requires that it has at least two roots with the same modulus.
This leads to the relation $H(|\beta(E|))=H(|\beta(E)|e^{i k})$ (here $k$ is a parameter), indicating that the aGBZ condition is also a degeneracy condition, meaning spectral self-crossing.
Based on our previous discussions, when there are crossing points in the spectra, the Wasserstein metric contains a term with finite-size scaling $\mathcal{O}(1/N)$.
In the limit $\mathrm{d}\mu\to0$, these finite-size terms exist when the exact degeneracy condition $H(|\beta|)=H(|\beta| e^{ik_j})$ is satisfied, with the phase different $k_j=2\pi j/N$ and $0\leq k_j\leq\pi$. Since there are $N/2$ possible values for $k_j$, the number of discrete points where the finite size effect occurs is $N/2$ for sufficiently small $\Delta\mu$.
To visualize these terms, we define $N\Delta G_W=N|G_W(\mu,N)-G_W(\mu,N\rightarrow \infty)|$, so that
the quantity $N\Delta G_W$ is independent of the system size $N$, and becomes finite at discrete values of $\mu$ satisfying the degeneracy condition $H(|\beta|)=H(|\beta| e^{ik_j})$, as illustrated in Fig.~\ref{Fig1}.

Because of the properties above, the Wasserstein metric can be used to determine the modulus range of the aGBZ and GBZ.
Specifically, for a given model, one can numerically evaluate the quantity $N\Delta G_W$, which typically features multiple regions in $\mu$ with non-vanishing $N\Delta G_W$. According to our previous discussions, these regions correspond to the modulus range of the aGBZs. Since the GBZ is the $p$th aGBZ, its modulus range can also be determined.

To illustrate the conclusions above, we now give two examples.
First, we calculate the Wasserstein metric for a general single-band Hamiltonian $H(\beta)=\sum_{n=-p}^q t_n \beta^n$, where
\begin{equation}
    G_W(\mu) = \sum_{n=-p}^{q}n^2|t_n|^2e^{2n\mu}.
\end{equation}
It is straightforward to apply the result above to HN model~\cite{hn}, whose Hamiltonian is given by
$H(\beta)=t_L\beta + t_R/\beta$, where $t_L$ and $t_R$ are the left- and right-ward hopping, respectively.
The Wasserstein metric is then
\begin{align}
    G_W(\mu) = 2t_Lt_R\cosh\left(2\mu + \ln\frac{t_L}{t_R}\right),
\end{align}
which possesses a minimum at $\mu_{\text{min}}=-\frac{1}{2}\ln\frac{t_L}{t_R}$.
Since the HN model has a circular GBZ, the radius of the GBZ is given by $\mu_{\text{min}}$.
In Fig.~\ref{Fig2}(a), we show the numerically evaluated metric.
Note that if we set $t_R=0$, which is the case for unidirectional hopping, the Wasserstein metric becomes $G_W(\mu) = t_L^2e^{2\mu}$, with its minimum ($G_W=0$) taking place at $\mu_{\text{GBZ}}=-\infty$.

We then consider the non-reciprocal SSH model~\cite{skin1}, with a next-nearest-neighbor hopping amplitude $t_3$. The Hamiltonian is given by
\begin{align}\label{SSH}
    H(\beta) & = \left[t_1+ \frac{1}{2}(t_2+t_3)(\beta+1/\beta)\right]\sigma_x \notag \\
    & + \left[ \frac{1}{2i}(t_2-t_3)(\beta-1/\beta) + \frac{i}{2}\gamma \right]\sigma_y,
\end{align}
where $\sigma_{x,y}$ are the Pauli matrices, $\gamma$ is the non-reciprocal parameter, and $t_1$ and $t_2$ are the nearest-neighbor hopping rates.

We first take $t_3=0$, and the Bloch Hamiltonian becomes
\begin{equation}\label{SSH2}
    H(\beta) = \begin{pmatrix}
        0 & t_1 + \gamma/2 + t_2\beta^{-1} \\
        t_1 - \gamma/2 + t_2\beta & 0
    \end{pmatrix}.
\end{equation}
While the GBZ is also circular in this case, the two-band nature of the model adds some complications. Specifically, as illustrated in Fig.~\ref{Fig2}(b), two singularities emerge in the Wasserstein metric, signaling the presence of EPs under the corresponding $\mu$.
They also give rise to multiple local minima.
Nevertheless, since the eigenspectrum
\begin{align}
    & E^2(k,\mu_{GBZ}+\mu)  =  t_1^2+t_2^2-\gamma^2/4+t_2\sqrt{|t_1^2-\gamma^2/4|}\notag \\
    & \times\big[\mathrm{sgn}(t_1+\gamma/2)e^{ik+\mu}+\mathrm{sgn}(t_1-\gamma/2)e^{-ik-\mu}\big].
\end{align}
remain unchanged when $\mu\to-\mu$, $k\to2\pi-k$ for $t_1>\gamma/2$, and $k\to \pi-k$ for $t_1<\gamma/2$, the Wasserstein metric
is also symmetric with respect to $\mu_{\text{GBZ}}$. This indicates that the central minimum corresponds to $\mu_{\text{GBZ}}$, yielding $\mu_{\text{GBZ}}=\ln\left|(t_1-\gamma/2)/(t_1+\gamma/2)\right|$.

Last, we consider the case with $t_3\neq 0$. We numerically evaluate the Wasserstein metric both in the thermodynamic limit (orange) and for a finite-size system (black), as illustrated in Fig.~\ref{Fig2}(c1)-(e1). Now that the GBZ is non-circular, the finite-size term in Eq.~(\ref{cross}) contributes in the aGBZs.
This is more apparent in Fig.~\ref{Fig2}(c2)-(e2), where the range of modulus for the aGBZs is determined as regions with finite $N\Delta G_W$.
The range of the GBZ is then determined by the $p$th region of finite $N\Delta G_W$ from the left,
as we show in Fig.~\ref{Fig2}(c2)-(e2).
Finally, in Fig.~\ref{Fig2}(c)-(e), we also show that the $\Delta G_W$ is indeed of the order $\mathcal{O}(1/N)$, since $N\Delta G_W$ remains largely unchanged with increasing $N$.

\subsection{Non-Bloch EPs}
   \begin{figure*}[tbp]
        \centering
        \includegraphics[width=1\linewidth]{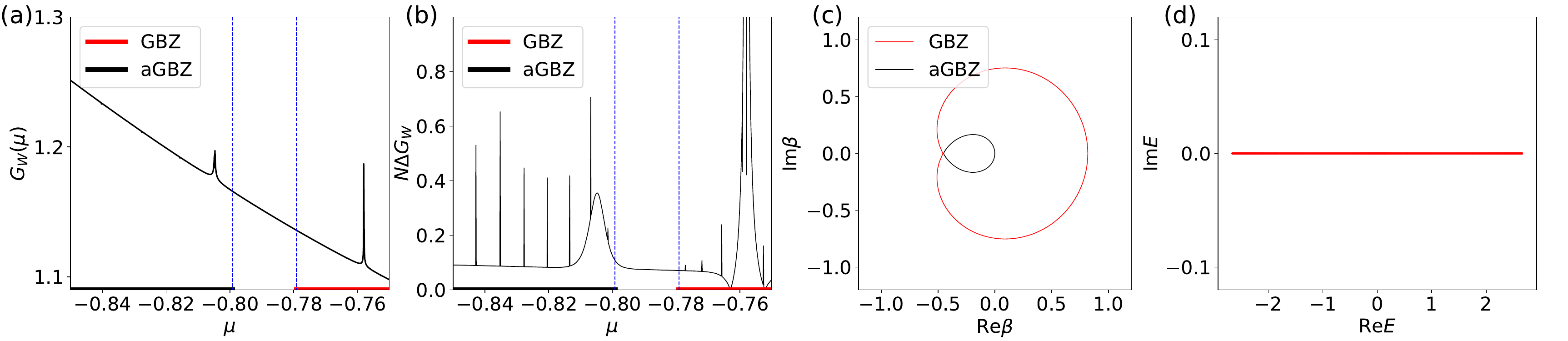}
        \caption{Non-Bloch EPs from the Wasserstein metric.
          (a)-(d) we investigate the non-Bloch EP, which occurs when the aGBZ merges with the GBZ. For a non-reciprocal SSH model, we set the hopping parameters to be $(t_1,t_2,t_3,\gamma)=(1.5609,1,0.2,4/3)$, the system size $N=400$ and $\Delta\mu=10^{-4}$.
          We fix $t_2,t_3,\gamma$ and change $t_1$, under which the non-Bloch EP lies at $t_1\simeq1.56$. In all subfigures, the modulus ranges of the aGBZ and GBZ are respectively indicated by the solid black and red lines on the $\mu$-axis.
        The blue vertical lines in (a)-(b) represent the boundaries of the (a)GBZ modulus ranges.}
        \label{Fig3}
    \end{figure*}
In this subsection, we discuss how the Wasserstein metric can be used to locate the non-Bloch EPs~\cite{pt1,pt2,pt3}.
In non-Hermitian systems, EPs can emerge only under the OBC, and visualized on the GBZ. These are the so-called non-Bloch EPs.

In Ref.~\cite{pt1}, it is shown that non-Bloch EPs generally emerge when the GBZ is not a smooth curve or when there are cusps on the GBZ. Specifically, at the non-Bloch EP transition point, the aGBZ merges with the GBZ, and the saddle points on the aGBZ and simultaneously become saddle points and cusps on the GBZ. This phenomenon is identified as the geometric origin of non-Bloch EPs.

Importantly, in describing a non-Bloch EP transition, this condition is equivalent to requiring that the modulus of the aGBZ and GBZ be equal at some point. That is to say, if two saddle points on the aGBZ and GBZ have the same modulus and the corresponding eigenvalue $E_s$, their phases must also be the same.
To prove this, we denote the saddle point on the GBZ as $\beta_{s,1}(E_s)$, which is a double root of the characteristic equation at $E_s$. According to the GBZ condition $|\beta_{p}(E_s)|=|\beta_{p+1}(E_s)|$, we identify $\beta_{s,1}(E_s)$ (which is on the GBZ) with either $\beta_p(E_s)$ or $\beta_{p+1}(E_s)$, leading to $\beta_p(E_s)=\beta_{p+1}(E_s)=\beta_{s,1}(E_s)$.
Similarly, denoting the saddle point on an adjacent aGBZ (to the GBZ) as $\beta_{s,2}(E_s)$, we have $\beta_{p-1}(E_s)=\beta_{p}(E_s)=\beta_{s,2}(E_s)$ or $\beta_{p+2}(E_s)=\beta_{p+1}(E_s)=\beta_{s,2}(E_s)$, which leads to our conclusion $\beta_{s,1}(E_s)=\beta_{s,2}(E_s)$.

The above statement indicates that when considering the non-Bloch EPs, we need only focus on the modulus of the (a)GBZ. Hence the condition for non-Bloch EPs can be relaxed to the requirement that the modulus range of the GBZ and that of its adjacent aGBZ just touches, a condition that can be determined by calculating $N\Delta G_W$, as discussed previously.

To support our conclusions above, we consider the non-reciprocal SSH model with $t_3\neq0$. Both the GBZ and aGBZ of the model are non-circular for $t_3\neq 0$. In Fig.~\ref{Fig3}(a)-(d), we can see clearly the range of modulus for the aGBZ (black) and GBZ (red) and they are going to merge. Note that the finite size effect in Fig.~\ref{Fig3}(a) is hard to observe due to a relatively large $N$. It is visible in Fig.~\ref{Fig3}(b) through the calculation of $N\Delta G_W$. Importantly, the non-Bloch EP occurs when the aGBZ touches the GBZ in (c), corresponding to the merging of the two regions with the finite-size contributions.

\subsection{EPs on the GBZ and topological transition}
In previous sections, we have demonstrated that singularities in the Wasserstein metric correspond to EPs for the non-Hermitian model under the corresponding imaginary flux $\mu$. In this subsection, we discuss its applications in predicting non-Bloch topological transitions.

In Ref.~\cite{semi1}, Yokomizo \textit{et al.} investigate systems with both sub-lattice symmetry and time-reversal symmetry and track the movement of EPs and suggest that the change of the winding number in this models can be understood by the motions of EPs on the GBZ. While EPs on the GBZ have richer implications~\cite{semi1,semi2,semi3}, we focus on detecting the change of winding  number (or a topological transition) through the Wasserstein metric.

Without loss of generality, we consider a two-band model
\begin{equation}
    H(\beta) = \begin{pmatrix}
        0 & R_+(\beta) \\
        R_-(\beta) & 0
    \end{pmatrix},
\end{equation}
where $R_\pm(\beta)=\sum_{n=-p_{\pm}}^{q_{\pm}} t_{\pm,n} \beta^n$. The non-Bloch winding number is~\cite{skin1,semi1}
\begin{align}
    w & = \frac{i}{4\pi}\int_{GBZ}\left(\frac{\mathrm{d}R_+}{R_+} - \frac{\mathrm{d}R_-}{R_-}\right) \notag \\
    & = -\frac{N_+-N_-}{2}+\frac{p_+-p_-}{2},
\end{align}
where $N_\pm$ denote the number of zeros of $\beta^{p_\pm}R_\pm(\beta)$ inside the GBZ. Note that EPs in this model arise whenever $R_\pm(\beta_s)=0$. If these points are further on the GBZ, they are referred to as EPs on the GBZ.

The change of winding number results from the change of $N_\pm$, which can be determined by the creation and annihilation of EPs with the same modulus on the GBZ~\cite{semi1}.
Hence, the topological transition can be deduced by tracking the movement of EPs on the GBZ. Since the Wasserstein metric contains the information of the modulus of GBZ and EPs, the above argument, translated to the Wasserstein metric, means tracking the movement of singularities on the GBZ.

\begin{figure}[tbp]
    \centering
    \includegraphics[width=\linewidth]{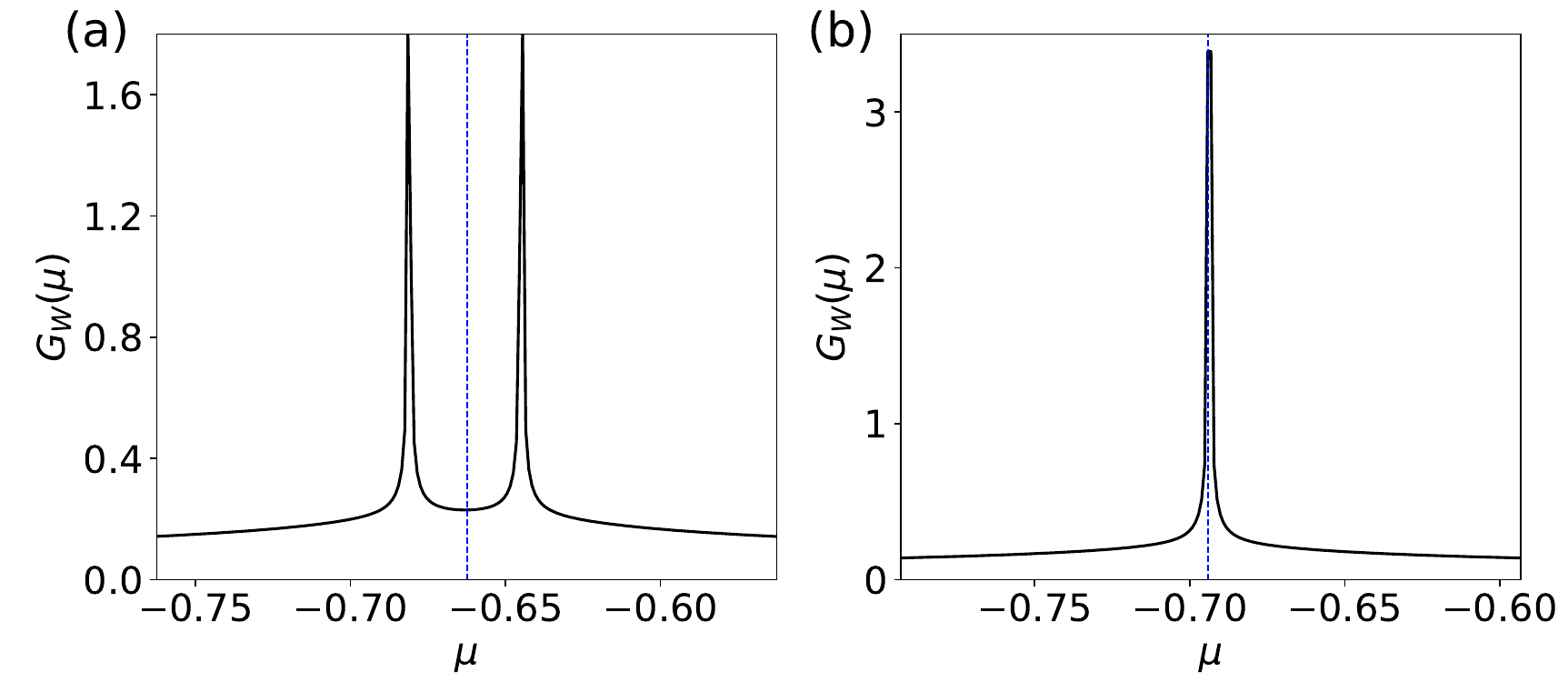}
    \caption{Movements of singularities of the Wasserstein metric on the GBZ. Here we set the hopping parameters of the non-reciprocal SSH model to be $t_2=0.4,\gamma=1$. According to $t_{1,c} = \pm\sqrt{\pm t_2^2 + (\gamma/2)^2}$, a topological transition occurs when $t_1 = \sqrt{-t_2^2 + (\gamma/2)^2}=0.30$. We set $t_1=0.29,0.30$ in (a) and (b), respectively. When the topological transition occurs, the two singularities merge to become two-fold degenerate.}
    \label{Fig4}
\end{figure}

To demosntrate this, we use the non-reciprocal SSH model in Eq.~(\ref{SSH}) as an example.
Setting $t_3=0$, we have a circular GBZ, with a radius $\sqrt{\left|(t_1-\gamma/2)/(t_1+\gamma/2)\right|}$. The non-Bloch topological transition occurs when two EPs on the GBZ have the same modulus, corresponding to the merging of two Wasserstein-metric singularities on the GBZ. This is given by
\begin{align}
    t_1 + \gamma/2 + t_2\beta_1^{-1} = t_1 - \gamma/2 + t_2\beta_2 = 0,
\end{align}
where we require $|\beta_1| = |\beta_2|$. The solutions to these equations are
\begin{align}
    t_{1,c} & = \pm\sqrt{\pm t_2^2 + (\gamma/2)^2},\\
    |\beta_1| & =|\beta_2| = \sqrt{\left|\frac{t_1-\gamma/2}{t_1+\gamma/2}\right|}.
\end{align}
The second expression above gives the non-Bloch topological transition point. Our numerical results are shown in Fig.~\ref{Fig4}. We observe the movement and merging of two singular points on the GBZ, resulting in the change of the winding number by 1. Therefore, the Wasserstein metric is also a useful tool to characterize the non-Bloch topological transition.

\section{Systems without translational symmetry}

\begin{figure*}[tbp]
    \centering
    \includegraphics[width=\linewidth]{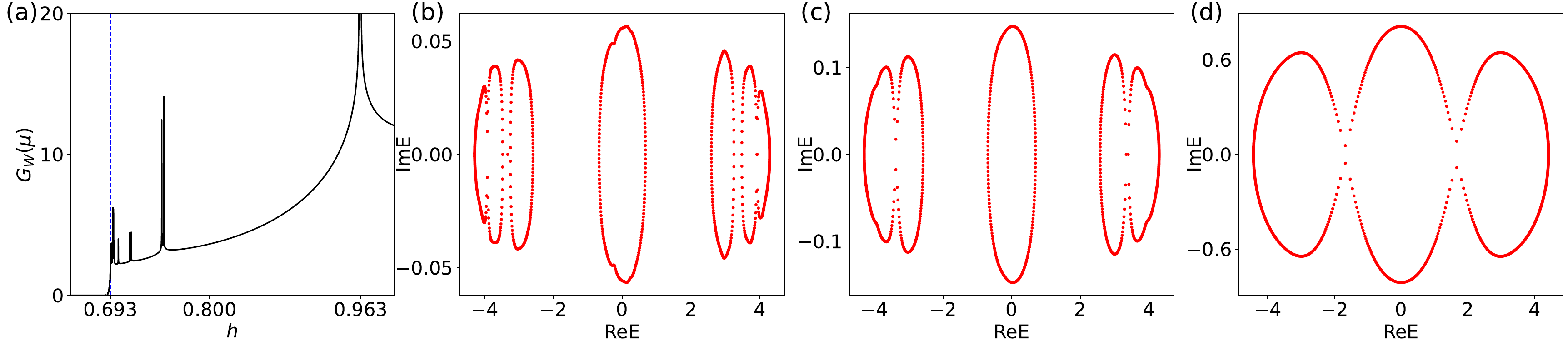}
    \caption{(a)The Wasserstein metric as a function of $h$ for the non-Hermitian AA model for a system size $N=1000$. A topological transition occurs at $h_c = \ln 2 \simeq 0.693$, where singularities (that is EPs) start to emerge. Since the eigenspectra of Hermitian AA model forms a Cantor set (for $h<h_c$), there are infinite number of singularities (EPs) for $h>h_c$ (see main text for discussion). The blue dotted vertical line marks the topological transition point, where the winding number $w_\phi$ changes. (b)-(d) Eigenspectra for (b) $h=0.715$, (c) $h=0.749$, and (d) $h=0.963$. The line gaps close at $h\simeq0.715,0.749,0.963$, which are also EPs. }
    \label{Fig5}
\end{figure*}
In previous sections, we discuss the utility of the Wasserstein metric in non-Hermitian models with lattice translational symmetry, where the concepts of the GBZ and aGBZ are available.
But the Wasserstein metric, and the idea behind it, are also applicable in systems without the translational symmetry, and capable of characterizing the rich spectral features therein. In this section, we show that the metric is also useful in predicting phase transitions in non-Hermitian quasiperiodic models~\cite{phase,quasi1,quasi2,quasi3,quasi4,quasi5,quasi6,quasi7,quasi8,quasi9,quasi10,quasi11,quasi12}.

We first consider a general non-Hermitian quasiperiodic model, characterized by both a complex potential and nonreciprocal hopping
\begin{align}\label{disorder}
     H(\theta, g) = &\sum_j^N(te^{-g}\ket{j}\bra{j+1} \notag \\
    & +te^g\ket{j+1}\bra{j}+V_j(\theta)\ket{j}\bra{j}).
\end{align}
Here the on-site potential $V_j(\theta)=\sum_{l=1}^d2\lambda_l\cos[l(2\pi\omega j+\theta)]$ and $\theta=\phi + ih$. We set $\omega=(\sqrt{5}-1)/2$ and $t=1$ as the unit of energy. We note that the dual Hamiltonian of Eq.~(\ref{disorder}) is
\begin{align}\label{dual}
    H' = & \sum_j \bigg[ \sum_{n=1}^d \big(\lambda_n e^{n(i\phi+h)}\ket{j+n} \bra{j} \notag \\
    & +\lambda_n e^{-n(i\phi+h)}\ket{j} \bra{j+n} \big)\notag \\
    & + 2 \cos(2\pi \omega j + i g)\ket{j}\bra{j} \bigg].
\end{align}

A key observation here is that we can interpret the imaginary phase $h$ as the imaginary flux $\mu$.
It follows that Eq.~(\ref{dual}) is a 1D non-Hermitian single-band model with an on-site potential. Since the dual transformation does not alter the spectrum, we analyze the Wasserstein metric in the $h$-space, in the hope of getting important information of the original model (\ref{disorder}).

Following Ref.~\cite{phase,quasi1,quasi2}, we define a winding number of the system according to
\begin{align}
     w_\phi(E_B,h,g) & = \lim_{N\to\infty}\frac{1}{2\pi i}\int_0^{2\pi}\mathrm{d}\phi \partial_\phi \frac{\ln\det[H(\theta, g)-E_B]}{N} \notag \\
    & = \lim_{N\to\infty}\frac{1}{2\pi i}\int_0^{2\pi}\mathrm{d}\phi \partial_\phi \zeta(E_B, \theta, g),
\end{align}
where $\zeta(E_B,\theta,g)=\left[\ln\det|H(\theta, g)-E_B|\right]/N$, with $E_B$ being the reference energy. Here the winding number typically characterizes a topological transition of the model~\cite{quasi1,quasi2}.

Analytically calculating the winding number, according to the Appendix in Ref.~\cite{quasi1}, we have
\begin{equation}
    \frac{1}{2\pi i}\partial_\phi\zeta(E_B, \theta, g) = \begin{cases} 0 & |g| > \gamma \\ \Omega(\phi) &  0<|g| < \gamma  \end{cases},
\end{equation}
where $\Omega(\phi)$ satisfies
\begin{align}
    \lim_{N\to\infty}\int_0^{2\pi}\Omega(\phi)\mathrm{d}\phi &= 2\pi w_\phi(E_B,h,0) i\nonumber\\
    &=-2\pi i\partial_h\gamma(E_B,h),
\end{align}
and $\gamma(E_B,h)$ is the Lyapunov exponent.
Notably, the Lyapunov exponent gives the inverse localization length of the system in the neighborhood of $E_B$~\cite{LE1,quasi1}, and can be calculated through the transfer matrix~\cite{LE2}.

We then obtain the winding number
\begin{align}
    & w_\phi(E_B,h,g) \notag \\
    & = \begin{cases} 0 & |g| > \gamma \\  w_\phi(E_B,h,0)=-\partial_h\gamma(E_B,h) &  0<|g| < \gamma  \end{cases}.
\end{align}
According to Ref.~\cite{quasi6}, the spectrum remains invariant for varying $h$, when all the eigenstates of Eq.(\ref{disorder}) are extended. This condition is equivalent to a vanishing
winding number for $|g|>\gamma$. Since the Wasserstein metric measures the change of the spectrum, the Wasserstein metric in the $h$-space remains zero when the system is topologically trivial with a vanishing winding number.

In addition, for finite systems, in the absence of spectral self-crossing, we also have $G_W(h) = \frac{1}{N}\sum_{i=1}^N\left|\frac{\partial E_i(\theta)}{\partial h}\right|^2$. Similar to Eq.~(\ref{EP}), the metric diverges when EPs occur anywhere in the parameter space.

Finally, we use the non-Hermitian AA model without non-reciprocal hopping~\cite{quasi1,quasi2,quasi8,quasi9} as a concrete example.
Since its spectrum forms a Cantor set~\cite{quasi1,quasi2,fractal} for $h = 0$, there are infinite line gaps in the spectrum. Numerically, we find that the line-gap closing points (where discrete spectral loops merge) are also EPs of the Hamiltonian. This is consistent with the results in
Ref.~\cite{quasi1,quasi2}, where only the single EP at the topological transition point is studied.
The Hamiltonian of the AA model is
\begin{align}
    H  = \sum_n^N&(\ket{n}\bra{n+1}+\ket{n+1}\bra{n} \notag \\
     & +2\lambda\cos(2\pi\omega n + \phi + ih)\ket{n}\bra{n}).
\end{align}
Here the Lyapunov exponnet is $\gamma=\ln|\lambda| + |h|$. When $|h|<|h_c|=|g|-\ln|\lambda|$, the system is topologically trivial and its spectrum remains unchanged as $h$ varies. As shown in Fig~.\ref{Fig5}, we numerically calculate the Wasserstein metric in the thermodynamic limit.
For $h<h_c$, both the winding number and the Wasserstein metric remain zero.
For $h>h_c$,  due to the fractal nature of the spectrum in the topological trivially regime, there are infinite line gaps and hence EPs.
This is reflected in the multitude of singularities in the Wasserstein metric for $h>h_c$. In Fig.~\ref{Fig5}(b)-(d), we show the eigenspectra for $h=0.715, 0.749,0.963$, respectively, where the line gaps close. This leads to divergence in the Wasserstein metric, indicating the sequential emergence of EPs with increasing $h$ [see Fig.~\ref{Fig5}(a)].

\section{Summary}
To summarize, we show that the optimal spectral transport, characterized by the Wasserstein metric, contains key spectral information of 1D non-Hermitian models. Since spectral geometry and topology play an important role in non-Hermitian models, the Wasserstein metric provides a shortcut to important features of the system, such as the geometry of the (a)GBZ, the non-Bloch EP and the topological transitions.
It turns out that the modulus of the (a)GBZ, which is reflected in the Wasserstein metric, offers sufficient information for determining a variety of spectral properties.

The Wasserstein metric we study here focuses on the spectral geometry in the $\mu$-space (which can be glossed as the imaginary $k$-space). This is in contrast to previous studies of quantum geometry of Hermitian bands in the real $k$-space ~\cite{qg1,qg2,qg3}. The quantum geometry depicts the structure of Bloch states in the Hilbert space, which are important for systems at equilibrium. Whereas by characterizing spectral geometry, the Wasserstein metric offers insight to the dynamic processes in non-Hermitian systems, which are intrinsically non-equilibrium.

For future studies, it would be interesting to explore the optimal transport problem and utility of the Wasserstein metric in non-Hermitian models of higher dimensions. Therein, the manifold of the imaginary-flux space can exhibit intriguing structures characterized by geodesics and curvatures. Moreover, the degeneracy condition for the PBC spectra in higher dimensions is much more complicated, making the correspondence between the Wasserstein metric and (a)GBZ in higher dimensions a nontrivial open question.

\begin{acknowledgments}
This work is supported by the National Natural Science Foundation of China (Grant No. 12374479), and by the Innovation Program for Quantum Science and Technology (Grant No. 2021ZD0301205). Z.G. acknowledges support from the University of Tokyo Excellent Young Researcher Program and from JST ERATO Grant Number JPMJER2302, Japan.
\end{acknowledgments}

\end{document}